\documentstyle[12pt]{article}
\input amssym.def
\input amssym.tex

\newcommand{\wx}{\widetilde{x}}
\newcommand{\wip}{\widetilde{p}}
\newcommand{\wa}{\widetilde{a}}
\newcommand{\wb}{\widetilde{b}}
\newcommand{\wc}{\widetilde{c}}
\newcommand{\wid}{\widetilde{d}}

\newcommand{\qed}{\rule{3mm}{3mm}}

\begin{document}

\def\m@th{\mathsurround=0pt}
\mathchardef\bracell="0365 
\def\upbrall{$\m@th\bracell$}
\def\undertilde#1{\mathop{\vtop{\ialign{##\crcr
    $\hfil\displaystyle{#1}\hfil$\crcr
     \noalign
     {\kern1.5pt\nointerlineskip}
     \upbrall\crcr\noalign{\kern1pt
   }}}}\limits}

\begin{titlepage}{\LARGE
\begin{center} Some further curiousities\\
from the world of integrable lattice systems \\
and their discretizations
 \end{center}} \vspace{1.5cm}
\begin{flushleft}{\large Yuri B. SURIS}\end{flushleft} \vspace{1.0cm}
Centre for Complex Systems and Visualization, University of Bremen,\\
Kitzb\"uhler Str. 2, 28359 Bremen, Germany\\
e-mail: suris @ cevis.uni-bremen.de

\vspace{1.5cm}  
{\small {\bf Abstract.} Unexpected relations are found between the Toda
lattice, the relativistic Toda lattice and the Bruschi--Ragnisco lattice,
as well as between their integrable discretizations.}
 
\end{titlepage}

\setcounter{equation}{0}
\section{Introduction}

We continue in this paper playing some curious game named ''reparametrizations''
with the well--known integrable lattice equations of the classical mechanics.
This game was started in the papers \cite{SNTL}, \cite{SNRTL}, where several
''new'' integrable lattices were introduced together with their integrable
discretizations.

In the present paper we shall demonstrate that the equations of motion of one
of simplest flow of the relativistic Toda hierarchy may be brought into the
form:
\begin{equation}\label{newrt}
\ddot{x}_k=\dot{x}_k\left(x_{k+1}-2x_k+x_{k-1}\right)+
\left(\frac{\dot{x}_{k+1}\dot{x}_k}{x_{k+1}-x_k}-
\frac{\dot{x}_k\dot{x}_{k-1}}{x_k-x_{k-1}}\right).
\end{equation}

This form of equations of motion is very remarkable, since the right--hand
side may be seen as a {\it sum} of the right--hand sides of another two 
well--known integrable lattices. The first one of them is the usual Toda 
lattice, the equations of motion for which can be put into the form:
\begin{equation}\label{newt}
\ddot{x}_k=\dot{x}_k\left(x_{k+1}-2x_k+x_{k-1}\right).
\end{equation}

The second one is exactly the so called Bruschi--Ragnisco lattice \cite{BR},
\cite{SBR}:
\begin{equation}\label{br}
\ddot{x}_k=\frac{\dot{x}_{k+1}\dot{x}_k}{x_{k+1}-x_k}-
\frac{\dot{x}_k\dot{x}_{k-1}}{x_k-x_{k-1}}.
\end{equation}

Even more bizarre becomes the situation when looking at the discrete time
counterparts of these systems. Namely, we shall demonstrate that the
discretization of the relativistic Toda flow derived in \cite{DRTL} in
the parametrization corresponding to (\ref{newrt}) takes the form

\begin{equation}\label{newdrt}
\frac{\wx_k-x_k}{x_k-\undertilde{x_k}}=
\frac{(x_{k+1}-x_k)}{(\undertilde{x_{k+1}}-x_k)}\;
\frac{(x_k-\wx_{k-1})}{(x_k-x_{k-1})}\;
\frac{\Big(1+h(\undertilde{x_{k+1}}-x_k)\Big)}{\Big(1+h(x_k-\wx_{k-1})\Big)}.
\end{equation}

The right--hand side of this equation may be seen as a {\it product} of
the right--hand sides of the following two equations:
\begin{equation}\label{newdt}
\frac{\wx_k-x_k}{x_k-\undertilde{x_k}}=
\frac{1+h(\undertilde{x_{k+1}}-x_k)}{1+h(x_k-\wx_{k-1})},
\end{equation}
and
\begin{equation}\label{dbr}
\frac{\wx_k-x_k}{x_k-\undertilde{x_k}}=
\frac{(x_{k+1}-x_k)}{(\undertilde{x_{k+1}}-x_k)}\;
\frac{(x_k-\wx_{k-1})}{(x_k-x_{k-1})}.
\end{equation}

Here (\ref{newdt}) turns out to be exactly the same discretization of the 
Toda lattice as derived in \cite{PGR}, \cite{DTL} and discussed further in 
\cite{SNTL},
but expressed in the parametriztion leading to (\ref{newt}). And (\ref{dbr})
is an integrable discretization of the Bruschi--Ragnisco lattice which, to
the author's knowledge, has not yet appeared in the literature, and can
be derived on the basis of a general recipe worked out in \cite{DTL},
\cite{DRTL} and the subsequent publications of the author.

All the systems above (continuous and discrete time ones) may be considered
either on an infinite lattice ($k\in{\Bbb Z}$), or on a finite one 
($1\le k\le N$). In the last case one of the two types of boundary conditions 
may be imposed: open--end ($x_0=\infty$, $x_{N+1}=-\infty$) or periodic
($x_0\equiv x_N$, $x_{N+1}\equiv x_1$). We shall be concerned only with the
finite lattices here, consideration of the infinite ones being to a large
extent similar.

\setcounter{equation}{0}
\section{Simplest flow of the Toda hierarchy}

We consider in this section the simplest flow of the Toda  hierarchy. 
All the results here are not new but collected in the form convenient 
for our present purposes. For the relevant references see \cite{DTL}.

The simplest flow of the Toda hierarchy is:
\begin{equation}\label{t}
\dot{a}_k=a_k(b_{k+1}-b_k), \quad \dot{b}_k=a_k-a_{k-1}.
\end{equation}
It may be considered either under open--end boundary conditions
($a_0=a_N=0$), or under periodic ones (all the subscripts are
taken (mod $N$), so that $a_0\equiv a_N$, $b_{N+1}\equiv b_1$). 

The flow (\ref{t}) is Hamiltonian with respect to the following Poisson  
bracket:
\begin{equation}\label{t l br}
\{a_k,b_k\}=-\{a_k,b_{k+1}\}=a_k
\end{equation}
(only the non--vanishing brackets are written down), with the Hamiltonian 
function
\begin{equation}\label{t H}
H_{\rm T}=\frac{1}{2}\sum_{k=1}^Nb_k^2+\sum_{k=1}^Na_k.
\end{equation}

An integrable discretization of the flow (\ref{t}) is given by the difference 
equations \cite{DTL}
\begin{equation}\label{dt}
\wa_k=a_k\frac{\beta_{k+1}}{\beta_k},\quad
\wb_k=b_k+h\left(\frac{a_k}{\beta_k}-\frac{a_{k-1}}{\beta_{k-1}}\right), 
\end{equation}
where $\beta_k=\beta_k(a,b)$ is defined as a unique set of functions 
satisfying the recurrent relation
\begin{equation}\label{t recur}
\beta_k=1+hb_k-\frac{h^2a_{k-1}}{\beta_{k-1}}
\end{equation}
together with an asymptotic relation
\begin{equation}\label{as beta}
\beta_k=1+hb_k+O(h^2).
\end{equation}
In the open--end case, due to $a_0=0$, we obtain from (\ref{t recur}) the 
following finite continued fractions expressions for $\beta_k$:
\[
\beta_1=1+hb_1;\quad 
\beta_2=1+hb_2-\frac{h^2a_1}{1+hb_1};\quad\ldots\quad;
\]
\[
\beta_N=1+hb_N-\frac{h^2a_{N-1}}{1+hb_{N-1}-
\displaystyle\frac{h^2a_{N-2}}{1+hb_{N-2}-
\parbox[t]{1.0cm}{$\begin{array}{c}\\  \ddots\end{array}$}
\parbox[t]{2.2cm}{$\begin{array}{c}
 \\  \\-\displaystyle\frac{h^2a_1}{1+hb_1}\end{array}$}}}.
\]
In the periodic case  (\ref{t recur}), (\ref{as beta}) uniquely define 
$\beta_k$'s as $N$-periodic infinite continued fractions. It can be 
proved that for $h$ small enough these continued fractions converge and their 
values satisfy (\ref{as beta}).

It can be proved \cite{DTL} that the map (\ref{dt}) is Poisson with respect 
to the bracket (\ref{t l br}).

Let us recall also the Lax representations of the flow (\ref{t}) and of the
map (\ref{dt}). They are given in terms of the  $N\times N$ Lax matrix $T$
depending on the phase space coordinates 
$a_k, b_k$ and (in the periodic case) on the additional parameter $\lambda$:
\begin{equation}\label{T}
T(a,b,\lambda) = \sum_{k=1}^N b_kE_{kk}+\lambda\sum_{k=1}^N E_{k+1,k}+
\lambda^{-1}\sum_{k=1}^N a_kE_{k,k+1}.
\end{equation}
Here $E_{jk}$ stands for the matrix whose only nonzero entry on the intersection
of the $j$th row and the $k$th column is equal to 1. In the periodic case we
have $E_{N+1,N}=E_{1,N}, E_{N,N+1}=E_{N,1}$; in the open--end case we set
$\lambda=1$, and $E_{N+1,N}=E_{N,N+1}=0$. 

The flow (\ref{t}) is equivalent to the matrix differential equation
\begin{equation}\label{t Lax}
\dot{T}=\left[ T,B\right], 
\end{equation}
where
\begin{equation}\label{B}
B(a,b,\lambda) = \sum_{k=1}^Nb_kE_{kk}+\lambda\sum_{k=1}^N E_{k+1,k},
\end{equation}
and the map (\ref{dt}) is equivalent to the matrix difference equation
\begin{equation}\label{dt Lax}
\widetilde{T}={\rm\bf B}^{-1}T{\rm\bf B},
\end{equation}
where
\begin{equation}\label{bB}
{\rm\bf B}(a,b,\lambda)=\sum_{k=1}^N\beta_kE_{kk}
+h\lambda\sum_{k=1}^NE_{k+1,k}.
\end{equation}

The spectral invariants of the matrix $T(a,b,\lambda)$ serve as 
integrals of motion for the flow (\ref{t}), as well as for the map
(\ref{dt}). In particular, 
\[
H_{\rm T}=\frac{1}{2}{\rm tr}(T^2).
\]

An involutivity of the spectral invariants of the matrix $T$   
follows from the $r$--matrix theory.
Namely, it turns out that the matrices $T(a,b,\lambda)$ form a Poisson
submanifold for a certain linear $r$--matrix bracket on the algebra of 
$N\times N$ matrices (in the open--end case) or on the loop algebra over
the previous one (in the periodic case) (cf. \cite{RS}), and (\ref{t l br})
gives exactly the corrdinate representation of this $r$--matrix bracket
restricted to the set of the matrices $T(a,b,\lambda)$.

\setcounter{equation}{0}
\section{Continuous and discrete time Toda flows\newline
in a new parametrization}

The usual parametrization of the variables $(a_k,b_k)$ by the canonically 
conjugated ones $(x_k,p_k)$ is:
\[
a_k=\exp(x_{k+1}-x_k),\quad b_k=p_k.
\]
This corresponds to the bracket (\ref{t l br}) and results in the most usual 
form of the Toda lattice and in its discretization studied in \cite{DTL}.

Here we chose a different parametrization, which, however, corresponds 
to the same bracket (\ref{t l br}) and is in a sense dual to the 
parametrization just mentioned:
\begin{equation}\label{new par t}
a_k=\exp(p_k),\quad b_k=x_k-x_{k-1}.
\end{equation}
The corresponding Hamiltonian function $H_{\rm T}$ takes the form
\begin{equation}
H_{\rm T}(x,p)=\sum_{k=1}^N\exp(p_k)+\frac{1}{2}\sum_{k=1}^N(x_k-x_{k-1})^2,
\end{equation}
which generates the canonical equations of motion
\begin{eqnarray*}
\dot{x}_k & = & \partial H_{\rm T}/\partial p_k=\exp(p_k),\\
\dot{p}_k & = & -\partial H_{\rm T}/\partial x_k=x_{k+1}-2x_k+x_{k-1}.
\end{eqnarray*}
These equations of motion imply the Newtonian ones (\ref{newt}).

We turn now to a less straightforward case of discrete equations of motion.

{\bf Theorem 1.} {\it In the parametrization} (\ref{new par t}) {\it the
equations of motion} (\ref{dt}) {\it may be presented in the form of the 
following two equations:
\begin{equation}\label{newdt:p}
h\exp(p_k) = (\wx_k-x_k)\Big(1+h(x_k-\wx_{k-1})\Big),
\end{equation}
\begin{equation}\label{newdt:wp}
h\exp(\wip_k) = (\wx_k-x_k)\Big(1+h(x_{k+1}-\wx_k)\Big),
\end{equation}
which imply also the Newtonian equations of motion} (\ref{newdt}).

{\bf Proof.} The second equation of motion in (\ref{dt}) together with the
second equation in (\ref{new par t}) implies that
\[
\wx_k-x_k-\frac{ha_k}{\beta_k}=\wx_{k-1}-x_{k-1}-\frac{ha_{k-1}}{\beta_{k-1}}.
\]
Hence the expression on the left--hand side does not depend on $k$, and setting
it equal to 0, we get:
\begin{equation}
\frac{ha_k}{\beta_k}=\wx_k-x_k.
\end{equation}
Substituting this into the recurrent relation (\ref{t recur}), we get:
\begin{equation}\label{beta}
\beta_k=1+h(x_k-\wx_{k-1}).
\end{equation}
The last two formulas imply the expression (\ref{newdt:p}) for $ha_k=h\exp(p_k)$.
Finally, the first equation of motion in (\ref{dt}) implies the expression
(\ref{newdt:wp}), if one takes into account (\ref{beta}). \qed

\setcounter{equation}{0}
\section{Simplest flow of the Bruschi--Ragnisco hierarchy}

This section is devoted to reminding the known facts about the simplest 
flow of the Bruschi--Ragnisco  hierarchy \cite{BR}, \cite{SBR}. 

It is given by the equations of motion:
\begin{equation}\label{rb}
\dot{b}_k=b_{k+1}c_k-b_kc_{k-1}, \quad \dot{c}_k=c_k(c_k-c_{k-1}).
\end{equation}
It may be considered either under open--end boundary conditions
($c_0=b_{N+1}=0$), or under periodic ones (all the subscripts are
taken (mod $N$), so that $c_0\equiv c_N$, $b_{N+1}\equiv b_1$). 

The flow (\ref{rb}) is Hamiltonian with respect to the following
Poisson bracket:
\begin{equation}\label{rb l br}
\{c_k,b_k\}=-\{c_k,b_{k+1}\}=c_k,
\end{equation}
with the Hamiltonian function 
\begin{equation}\label{rb H}
H_{\rm BR}=\sum_{k=1}^Nb_{k+1}c_k.
\end{equation}

The Lax representation of the flow (\ref{rb}) is given in terms of the  
$N\times N$  rank 1 Lax matrix $T(b,c)$  with the entries given by:
\begin{equation}\label{rb T}
T_{kj}=\left\{\begin{array}{l}b_j\prod_{i=k}^{j-1}c_i\;,\quad k\le j\\ \\
b_j\left(\prod_{i=j}^{k-1}c_i\right)^{-1},\quad k>j\end{array}\right.
\end{equation}
The flow (\ref{rb}) is equivalent to the following matrix differential equation:
\begin{equation}\label{rb Lax}
\dot{T}=\left[ T,M\right], 
\end{equation}
where $M$ is a {\it constant} matrix, so that the Lax equation is in fact
linear. For example, in the open--end case
\begin{equation}\label{M}
M = \sum_{k=1}^{N-1} E_{k+1,k}.
\end{equation}
Hence the solution of the flow (\ref{rb}) can be read off the following
explicit formula:
\begin{equation}
T(t)=\exp(-tM)T(0)\exp(tM).
\end{equation}

The integrals of motion for the flow (\ref{rb}) are given not by the spectral
invariants of the matrix $T(b,c)$ (they are all dependend because this matrix
has reank 1), but by the following set of linear functionals on the Lax matrix:
${\rm tr}(M^mT)$. In particular,
\[
H_{\rm BR}={\rm tr}(MT).
\]

The involutivity of such linear functionals on the matrix $T$ with respect to  
the bracket (\ref{rb l br}) follows from its interpretation as a coordinate 
representation of a standard Lie--Poisson bracket on the algebra $gl(N)$,
restricted to a Poisson submanifold of rank 1 matrices \cite{SBR}.
Taking such functions as Hamiltonians, i.e. replacing in the previous formulas 
$M$ by $M^m$, $1\le m\le N-1$, we get the whole Bruschi--Ragnisco hierarchy.

According to the philosphy of \cite{DTL}, \cite{DRTL}, the discrete time
Bruschi--Ragnisco lattice belongs to the same hierarchy, i.e. it shares 
the Lax matrix with the continuous time one, and its explicit solution
should be given by
\begin{equation}
T(nh)=(I+hM)^{-n}T(0)(I+hM)^n.
\end{equation}
Hence the corresponding discrete Lax equation should have the form
\begin{equation}\label{drb Lax}
\widetilde{T}=(I+hM)^{-1}T(I+hM).
\end{equation}

A direct calculation shows that the last equation in the coordinates 
$(b_k,c_k)$ has the form:
\begin{equation}\label{drb}
\wb_k(1+h\wc_{k-1})=b_k+hb_{k+1}c_k, \quad
\wc_k=c_k\;\frac{1+h\wc_k}{1+h\wc_{k-1}}. 
\end{equation}
By construction, this map is Poisson with respect to the bracket 
(\ref{rb l br}).

\setcounter{equation}{0}
\section{Continuous and discrete time Bruschi--Ragnisco \newline
lattices in the Newtonian form}

Observe now that the bracket (\ref{rb l br}) for the Bruschi--Ragnisco
lattice formally is identical with that for the Toda lattice (\ref{t l br}).

This suggest the parametrization
\begin{equation}\label{par rb}
c_k=\exp(p_k),\quad b_k=x_k-x_{k-1}.
\end{equation}
The corresponding Hamiltonian function $H_{\rm BR}$ takes the form
\begin{equation}
H_{\rm BR}(x,p)=\sum_{k=1}^N\exp(p_k)(x_{k+1}-x_k),
\end{equation}
which generates the canonical equations of motion
\begin{eqnarray*}
\dot{x}_k & = & \partial H_{\rm BR}/\partial p_k=\exp(p_k)(x_{k+1}-x_k),\\
\dot{p}_k & = & -\partial H_{\rm BR}/\partial x_k=\exp(p_k)-\exp(p_{k-1}).
\end{eqnarray*}
These equations of motion imply the Newtonian ones (\ref{br}).

Turning to a less straightforward case of discrete equations of motion, we have:

{\bf Theorem 2.} {\it In the parametrization} (\ref{par rb}) {\it the
equations of motion} (\ref{drb}) {\it may be presented in the form of the 
following two equations:
\begin{equation}\label{dbr:p}
h\exp(p_k) = \frac{(\wx_k-x_k)}{(x_k-\wx_{k-1})}\;
\frac{(x_k-x_{k-1})}{(x_{k+1}-x_k)},
\end{equation}
\begin{equation}\label{dbr:wp}
h\exp(\wip_k) = \frac{(\wx_k-x_k)}{(x_{k+1}-\wx_k)},
\end{equation}
which imply also the Newtonian equations of motion} (\ref{dbr}).

{\bf Proof.} We start with the chain of identities following from
the equations of motion (\ref{drb}), independently of any parametrization:
\[
b_{k+1}-\frac{b_{k+1}}{1+h\wc_k}=\frac{hb_{k+1}\wc_k}{1+h\wc_k}=
\frac{hb_{k+1}c_k}{1+h\wc_{k-1}}=\wb_k-\frac{b_k}{1+h\wc_{k-1}}.
\]
Now we substitute $b_{k+1}=x_{k+1}-x_k$ into the leftmost part of this chain,
$\wb_k=\wx_k-\wx_{k-1}$ into the rightmost part, and arrive at the conclusion
that
\[
\frac{b_{k+1}}{1+h\wc_k}-(x_{k+1}-\wx_k)=
\frac{b_k}{1+h\wc_{k-1}}-(x_k-\wx_{k-1}).
\]
So, the expression on the left hand side of this equality does not depend 
on $k$. Setting it equal to 0, we obtain:
\[
1+h\wc_k=\frac{b_{k+1}}{x_{k+1}-\wx_k}=\frac{x_{k+1}-x_k}{x_{k+1}-\wx_k},
\]
which is, clearly, equivalent to (\ref{dbr:wp}), and, together with
the second equation of motion in (\ref{drb}), implies also (\ref{dbr:p}). 
\qed

\setcounter{equation}{0}
\section{Simplest flow of the relativistic Toda hierarchy}

In this section we recall some known facts about the relativistic Toda 
lattice. The sources are: \cite{DRTL} and the references therein.

The simplest flow of the relativistic Toda hierarchy is:
\begin{equation}\label{rt}
\dot{d}_k=d_k(c_k-c_{k-1}), \quad
\dot{c}_k=c_k(d_{k+1}+c_{k+1}-d_k-c_{k-1}).
\end{equation}
It may be considered either under open--end boundary conditions
($d_{N+1}=c_0=c_N=0$), or under periodic ones (all the subscripts are
taken (mod $N$), so that $d_{N+1}\equiv d_1$, $c_0\equiv c_N$, 
$c_{N+1}\equiv c_1$). 

It is Hamiltonian with respect to the Poisson bracket
\begin{equation}\label{rt l br}
\{c_k,d_{k+1}\}=-c_k, \quad \{c_k,d_k\}=c_k,\quad \{d_k,d_{k+1}\}=c_k,
\end{equation}
with a Hamiltonian function 
\begin{equation}\label{rt H}
H_{\rm RT}=\frac{1}{2}\sum_{k=1}^N(d_k+c_{k-1})^2+\sum_{k=1}^N(d_k+c_{k-1})c_k.
\end{equation}

An integrable discretization of the flow (\ref{rt}) derived in \cite{DRTL} 
is given by the difference equations
\begin{equation}\label{drt}
\widetilde{d}_k=d_k\frac{{\goth a}_{k+1}-hd_{k+1}}{{\goth a}_k-hd_k},\quad
\widetilde{c}_k=c_k\frac{{\goth a}_{k+1}+hc_{k+1}}{{\goth a}_k+hc_k}, 
\end{equation}
where ${\goth a}_k={\goth a}_k(c,d)$ is defined as a unique set of functions 
satisfying the recurrent relation
\begin{equation}\label{rt recur}
{\goth a}_{k+1}=1+hd_{k+1}+\frac{hc_k}{{\goth a}_k}
\end{equation}
together with an asymptotic relation
\begin{equation}\label{as a}
{\goth a}_k=1+h(d_k+c_{k-1})+O(h^2).
\end{equation}
In the open--end case, due to $c_0=0$, one obtains from (\ref{rt recur}) the 
following finite continued fractions expressions for ${\goth a}_k$:
\[
{\goth a}_1=1+hd_1;\quad 
{\goth a}_2=1+hd_2+\frac{hc_1}{1+hd_1};\quad\ldots\quad;
\]
\[
{\goth a}_N=1+hd_N+\frac{hc_{N-1}}{1+hd_{N-1}+
\displaystyle\frac{hc_{N-2}}{1+hd_{N-2}+
\parbox[t]{1.0cm}{$\begin{array}{c}\\  \ddots\end{array}$}
\parbox[t]{2.2cm}{$\begin{array}{c}
 \\  \\+\displaystyle\frac{hc_1}{1+hd_1}\end{array}$}}}.
\]
In the periodic case  (\ref{rt recur}), (\ref{as a}) uniquely define 
${\goth a}_k$'s as $N$-periodic infinite continued fractions. It can be 
proved that for $h$ small enough these continued fractions converge and their 
values satisfy (\ref{as a}).

The map (\ref{drt}) is Poisson with respect to the bracket (\ref{rt l br}).

Both the continuous time flow (\ref{rt}) and the discrete time one 
(\ref{drt}) admit Lax representations with the same Lax matrix, given by
\begin{equation}
T(c,d,\lambda)=L(c,d,\lambda)U^{-1}(c,d,\lambda),
\end{equation}
where $L$ and $U$ are two $N\times N$ matrices depending on the phase space 
coordinates $c_k, d_k$ and (in the periodic case) on the additional parameter 
$\lambda$:
\begin{eqnarray}
L(c,d,\lambda) & = & \sum_{k=1}^N d_kE_{kk}+\lambda\sum_{k=1}^N E_{k+1,k},\\
U(c,d,\lambda) & = & \sum_{k=1}^N E_{kk}-\lambda^{-1}\sum_{k=1}^N 
c_kE_{k,k+1}.
\end{eqnarray}

The flow (\ref{rt}) is equivalent to the matrix differential equation
\begin{equation}
\dot{T}=\left[ T,A\right],
\end{equation}
where
\begin{equation}
A(c,d,\lambda)  =  \sum_{k=1}^N(d_k+c_{k-1})E_{kk}+\lambda
\sum_{k=1}^N E_{k+1,k}.
\end{equation}
The map (\ref{drt}) is equivalent to the matrix difference equation
\begin{equation}
\widetilde{T}={\rm\bf A}^{-1}T{\rm\bf A},
\end{equation}
where
\begin{equation}\label{bA}
{\rm\bf A}(c,d,\lambda)=\sum_{k=1}^N{\goth a}_kE_{kk}
+h\lambda\sum_{k=1}^NE_{k+1,k},
\end{equation}
and the quantities ${\goth a}_k$ are defined by the recurrent relations 
(\ref{rt recur}).

The spectral invariants of the matrices $T(c,d,\lambda)$ serve as integrals 
of motion for the flow (\ref{rt}), as well as for the map (\ref{drt}). 
In particular,
\[
H_{\rm RT}=\frac{1}{2}{\rm tr}(T^2).
\]

As in the case of the Toda lattice, the involutivity of the spectral invariants 
of the matrix $T$ with respect to the bracket (\ref{rt l br})
follows from the $r$--matrix theory.
Namely, it turns out that the matrices $T(c,d,\lambda)$ form another Poisson
submanifold for the same bracket on the matrix algebra as in the Toda case,
and  (\ref{rt l br}) serves as a coordinate representation of this bracket
restricted to this manifold (cf. \cite{S2}).

\setcounter{equation}{0}
\section{Continuous and discrete time relativistic Toda \newline
lattices in a new parametrization}

It is easy to see that the following parametrization of the variables $(c_k,d_k)$
by the canonically conjugated variables $(x_k,p_k)$ leads to the linear bracket
(\ref{rt l br}):
\begin{equation}\label{new rt par}
d_k=x_k-x_{k-1}-\exp(p_{k-1}),\quad c_k=\exp(p_k).
\end{equation}

Let us find out how do the equations of motion look in this parametrization. We
start with (\ref{rt}).

Obviously, the function $H_{\rm RT}$ takes the form
\begin{equation}\label{rt H in xp}
H_{\rm RT}=\sum_{k=1}^N\exp(p_k)(x_k-x_{k-1})+
\frac{1}{2}\sum_{k=1}^N(x_k-x_{k-1})^2.
\end{equation}
Correspondingly, the flow (\ref{rt}) takes the form of canonical equations 
of motion:
\begin{eqnarray*}
\dot{x}_k & = & \partial H_{\rm RT}/\partial p_k=\exp(p_k)(x_k-x_{k-1}),\\
\dot{p}_k & = & -\partial H_{\rm RT}/\partial x_k=x_{k+1}-2x_k+x_{k-1}+
\exp(p_{k+1})-\exp(p_k).
\end{eqnarray*}
As an immediate consequence of these equations one gets the Newtonian equations
of motion (\ref{newrt}).

We turn now to a less straightforward case of discrete equations of motion.

{\bf Theorem 3.} {\it In the parametrization} (\ref{new rt par}) {\it the
equations of motion} (\ref{drt}) {\it may be presented in the form of the 
following two equations:
\begin{equation}\label{drt:p}
h\exp(p_k) = \frac{(\wx_k-x_k)}{(x_k-\wx_{k-1})}\;\Big(1+h(x_k-\wx_{k-1})\Big),
\end{equation}
\begin{equation}\label{drt:wp}
h\exp(\wip_k) = \frac{(\wx_k-x_k)}{(x_{k+1}-\wx_k)}\;
\frac{(\wx_{k+1}-\wx_k)}{(\wx_k-\wx_{k-1})}\;
\Big(1+h(x_{k+1}-\wx_k)\Big).
\end{equation}
which imply also the Newtonian equations of motion} (\ref{newdrt}).

{\bf Proof.} We start with derivation of several useful formulas which
do not depend on the parametrization of the variables $(c_k,d_k)$.

The equations of motion (\ref{drt}) together with the recurrent 
relation (\ref{rt recur}) imply:
\begin{equation}\label{aux1}
\wid_{k}+\wc_{k-1}=\left(d_k+c_{k-1}\,\frac{{\goth a}_k}{{\goth a}_{k-1}}\right)
\frac{{\goth a}_{k+1}-hd_{k+1}}{{\goth a}_k-hd_k}.
\end{equation}
But, applying once more (\ref{rt recur}), we derive:
\[
hd_k+hc_{k-1}\,\frac{{\goth a}_k}{{\goth a}_{k-1}}=
hd_k+{\goth a}_k({\goth a}_k-1-hd_k)=
({\goth a}_k-1)({\goth a}_k-hd_k),
\]
so that (\ref{aux1}) may be rewritten as
\begin{equation}\label{aux2}
h(\wid_{k}+\wc_{k-1})=({\goth a}_k-1)({\goth a}_{k+1}-hd_{k+1}).
\end{equation}
It is easy to see that, according to (\ref{rt recur}), 
\[
({\goth a}_k-1)({\goth a}_{k+1}-hd_{k+1})=
h(d_{k+1}+c_k)-({\goth a}_{k+1}-{\goth a}_k)\,,
\]
so that  (\ref{aux2}) implies:
\begin{equation}
{\goth a}_{k+1}-{\goth a}_k=h(d_{k+1}+c_k)-h(\wid_{k}+\wc_{k-1}).
\end{equation}

Now we are in a position to make use of the concrete parametrization
(\ref{new rt par}). Namely, due to $d_{k+1}+c_k=x_{k+1}-x_k$ the last formula
implies that
\[
{\goth a}_k-h(x_k-\wx_{k-1})={\goth a}_{k+1}-h(x_{k+1}-\wx_k),
\]
i.e. the expression on the left--hand side does not depend on $k$. 
Setting this equal to 1, we arrive at:
\begin{equation}\label{a in x}
{\goth a}_k=1+h(x_k-\wx_{k-1}).
\end{equation}
Now the formula (\ref{rt recur}) upon use of (\ref{new rt par}), 
(\ref{a in x}) may be presented as
\[
1+h(x_{k+1}-\wx_k)=1+h\Big(x_{k+1}-x_k-\exp(p_k)\Big)+
\frac{h\exp(p_k)}{1+h(x_k-\wx_{k-1})},
\]
which can be easily resolved for $\exp(p_k)$ and gives (\ref{drt:p}). Next,
(\ref{a in x}) together with the expression (\ref{drt:p}) for 
$c_k=\exp(p_k)$ implies
\[
{\goth a}_k+hc_k=\frac{(\wx_k-\wx_{k-1})}{(x_k-\wx_{k-1})}
\Big(1+h(x_k-\wx_{k-1})\Big).
\]
Substituting this into the second equation of motion in (\ref{drt}), 
we finally obtain (\ref{drt:wp}). \qed

\setcounter{equation}{0}
\section{Conclusion}

The relations between three well known lattice systems, established in
the present paper, may be symbolically presented as
\[
\rm RTL\;=\;TL\;+\;BRL\;, \qquad dRTL\;=\;dTL\;\times\;dBRL\;.
\]
They are, in my opinion, rather beautiful and unexpected. However,
they probably are nothing more than a pure curiosity, which demonstrates
once more: the field of integrable systems of classical mechanics, even 
in its best studied parts, is far from being exhausted. Many important 
and less important but funny findings await us in this fascinating
world.


\begin{thebibliography}{10}
\bibitem[1]{SNTL} Yu.B.Suris. A new integrable system related to the 
Toda lattice. -- solv-int/9605010.

\bibitem[2]{SNRTL} Yu.B.Suris. New integrable systems related to the 
relativistic Toda lattice. -- solv-int/9605006.

\bibitem[3]{BR} M.Bruschi, O.Ragnisco. A new integrable system with nearest
neighbours interaction. -- {\it Inverse Problems, 1989, v.5, p.983.}

\bibitem[4]{SBR} Yu.B.Suris. On the algebraic structure of the 
Bruschi--Ragnisco lattice. -- {\it Physics Letters A, 1993, v.179, p.403--406}.

\bibitem[5]{DRTL} Yu.B.Suris. A discrete--time relativistic Toda lattice. -- 
{\it J.Physics A: Math. \& Gen., 1996, v.29, p.451--465}.

\bibitem[6]{PGR} V.Papageorgiou, B.Grammaticos, A.Ramani. Orthogonal polynomial
approach to discrete Lax pairs for initial--boundary value problems of the $QD$
algorithm. -- {\it Letters Math. Phys, 1995, v.34.}

\bibitem[7]{DTL} Yu.B.Suris. Bi--Hamiltonian structure of the $qd$ algorithm
and new discretizations of the Toda lattice. -- 
{\it Physics Letters A, 1995, v.206, p.153--161}.

\bibitem[8]{RS} A.G.Reyman, M.A.Semenov-Tian-Shansky. Group--theoretical 
method in the theory of finite--dimensional integrable systems. --
{\it In: Encyclopaedia of Math.Sciences, V.16, Dynamical systems VII. 
Springer, 1993.}

\bibitem[9]{S2} Yu.B.Suris. On the bi--Hamiltonian structure of Toda and 
relativistic  Toda lattices. -- {\it Physics Letters A, 1993, v.180, 
p.419--429.}

\end{thebibliography}
\end{document}